\documentstyle[preprint,revtex]{aps}
\begin{document}
\draft
\preprint{Princeton,May12}
\preprint{May 1993}
\begin{title}
Solutions to the Multi-component $1/r$ Hubbard Model
\end{title}
\author{D. F. Wang$^\dagger$}
\begin{instit}
Joseph Henry Laboratories of Physics, Princeton University, \\
Princeton, New Jersey 08544
\end{instit}
\begin{abstract}
In this work we introduce one dimensional multi-component
Hubbard model of $1/r$ hopping and $U$ on-site energy.
The wavefunctions, the spectrum and the thermodynamics are
studied for this model in the strong interaction limit $U=\infty$.
In this limit, the system is a
special example of $SU(N)$ Luttinger liquids, exhibiting
spin-charge separation in the full Hilbert space.
Speculations on the physical properties of the model
at finite on-site energy are also discussed.
\end{abstract}

\narrowtext
Recent studies on the low dimensional systems have renewed great
interests in the Gutzwiller-Jastrow wavefunctions.
The wavefunctions are useful in the aspects that they may serve
as good variational wavefunctions or they may be exact solutions
of the Hamiltonians
\cite{girvin,arovas,mele,hald88,shas88,hald91,kura91,kawa91,Kawakami92,suth72,Rucken91,anderson89,note,ha92,wang92,wang,sutherland2,sutherland3,Frahm}.
Most recently, the intense research in the field of integrable systems
has shown the wavefunctions to be exact solutions of
some quantum many particle systems.
These integrable systems are characterized by that the full spectrum may even
be written in terms of more generalized Jastrow wavefunctions,
as in the cases of $1/r^2$ Fermi or Bose gases,
$1/r^2$ Haldane-Shastry spin chain, and the 1D supersymmetric t-J model of
$1/r^2$ hopping and exchange.

Hubbard model has been of great interest since the
discovery of the high Tc superconductivity.
About two years ago, Gebhard and Ruckenstein introduced
the one dimensional $SU(2)$ Hubbard model of $1/r$ hopping and
on-site energy $U$\cite{Rucken91}. The model is completely integrable
for arbitrary on-site energy.
In the strong interaction limit $U=\infty$, it has
been discovered recently that a set of Gutzwiller-Jastrow wavefunctions
is exact eigen-functions of the Hamiltonian\cite{wang},
the system exhibits the spin-charge separation in the full
Hilbert space\cite{Rucken91,wang}.

In this work, we introduce a new integrable model, the one
dimensional $1/r$ multi-component Hubbard model.
In the following we only discuss the strong interaction
limit $U=\infty$.
Generalizing our previous work, we show that a set
of $SU(N)$ Gutzwiller-Jastrow wavefunctions is eigenstates of the system.
The full excitation spectrum and the thermodynamics are also given
explicitly in this strong interaction case. Spin and charge
are decoupled in the full Hilbert space, the system is a special
example of $SU(N)$ Luttinger liquids.
At the end of the work, we also discuss speculations of further
investigation of the system of finite on-site energy $U$.


The Hamiltonian for the one-dimensional Hubbard model is given by
\begin{equation}
H=\sum_{\sigma}\sum_{i\ne j} t_{ij} c_{i\sigma}^{\dagger}
c_{j\sigma} + U \sum_{i} \sum_{\sigma \ne \sigma\prime}
n_{i\sigma} n_{i\sigma\prime},
\label{eq:hamil}
\end{equation}
where $c_{i\sigma}^{\dagger}$ and $c_{i\sigma}$ are creation and
annihilation operators at site $i$ with spin component $\sigma$.
The sum over $\sigma$ runs from
1 to $N$, where $N$ is the number of flavors of the fermions.
We take the hopping matrix $t_{ij}=i t (-1)^{(i-j)}/d(i-j)$
where $d(n)=(L/\pi)\sin(n\pi/L)$, and $U$ is the on-site energy.
Here, because of the special form of the  hopping matrix, for the
wavefunctions of the system, we assume
periodic boundary conditions for odd $L$, or anti-periodic boundary
condition for even $L$.

In the strong interaction limit $U=\infty$, each site can be occupied
at most by one particle. In this limit, we work in the Hilbert space of
no double occupancy and no multi-occupancy. The Hamiltonian can be
written in terms of the Hubbard operators:
\begin{equation}
H= \sum_{\sigma =1,2,\cdots,N} \sum_{i\ne j}
t_{ij} X_i^{\sigma 0 } X_j^{0\sigma}.
\end{equation}
Let us denote the number of holes by $Q$,
that of the fermions of the first flavor by $M_1$,
that of the second flavor by $M_2$, $\cdots$,
that of the $N-$th flavor by $M_N$.
Following notations used in previous literatures,
states in the Hilbert space can be represented by spin and hole
excitations from the state full of fermions with the $N-$th flavor
$|P\rangle$,
\begin{equation}
|\Phi> = \sum_{(\alpha,i), j} \Phi (\{x_i^\alpha\},\{y_j\})
\prod_{\alpha,i} b_{i \alpha}^{\dagger}
\prod_{j} h_j^{\dagger} |P>,
\end{equation}
Here $b_{i\alpha}^{\dagger} =c_{i\alpha}^{\dagger}c_{iN}$
annihilates one $N-th$ flavored fermion at site $i$ and
creats one $\alpha -th$ flavored fermion at site $i$ for
$\alpha =1, 2, \cdots, (N-1)$, while
$h_j^{\dagger} = c_{jN}$ creats a hole at site $j$. Here or in the
following we always
implicitly assume that we work in the space of no double
occupancy and no multi-occupancy in the discussion of the strong
interaction limit.
The amplitude $\Phi (\{x_i^\alpha\},\{y_j\})$ is symmetric
when exchanging the fermions at positions $x_i^{\alpha}$ and
$x_j^{\alpha}$, and antisymmetric
in the hole positions $\{y_i\}$.

Let us consider the following generalized $SU(N)$
Gutzwiller-Jastrow wavefunctions corresponding
to uniform motion and magnetization\cite{Kawakami92},
\begin{eqnarray}
&&\Phi_G (\{x_i^\alpha\},\{y_m\})= \exp {2\pi i \over L}[
\sum_{\alpha}J_{\alpha}\sum_i x_{i}^{\alpha}
+J_h \sum_i y_i] \times \Phi_0\nonumber\\
\Phi_0=\prod_{\alpha; i<j}&& d^2(x_i^{\alpha}
-x_j^{\alpha}) \cdot \prod_{\alpha<\beta; i,j} d(x_i^{\alpha}
-x_j^{\beta}) \cdot \prod_{\alpha,i,m} d(x_i^{\alpha}-y_m) \cdot
\prod_{m<n} d(y_m-y_n)
\end{eqnarray}
where $\alpha ,\beta = 1, 2, \cdots, N-1$. The quantum numbers $J_\alpha$ and
$J_h$ govern the momenta of the fermions and the holes.
They can be integers or half integers
such that the wavefunctions are periodic (or anti-periodic) for
odd $L$ ( for even $L$) under the translations
$x_i^{\alpha} \rightarrow x_i^{\alpha} +L$, and
$y_m \rightarrow y_m+L$. For the wavefunctions to be eigen-states of
the Hamiltonain, the quantum numbers must be choosen from some
restricted regions, which are to be specified below.


To demonstrate that the wavefunctions are eigen-states of the
Hamiltonian, we have to consider the
effect of the hopping operator very carefully.
The hopping operator can be broken into $N$ parts, each corresponding
to the hopping of fermions of different flavors.
Let us first consider the hopping operator of the $N-$th flavor,
${\hat T}(N) = \sum_{i\ne j} t_{ij} c_{iN}^{\dagger} c_{jN}$.
When it operates on the wavefunctions,
the hopping of the fermions of $N-$th flavor
is equivalent to the hopping of holes,
\begin{equation}
{{\hat T}(N) \Phi_G \over \Phi_G}
= - it \sum_{{\bar n} = 1}^{L-1} {(-1)^{\bar n} \over d({\bar n})}
z^{{\bar n} J_h} \sum_{n} \prod_{m\ne n}
F_{nm}({\bar n}) \prod_{(\alpha,i)} F_{n (\alpha,i)}({\bar n}),
\end{equation}
where $F_{nm}({\bar n}) = \cos{{\bar n} \pi \over L}
+\sin{{\bar n} \pi \over L} \cot \Theta_{nm},
F_{n(\alpha i)}({\bar n}) = \cos{{\bar n}\pi \over L}
+\sin{{\bar n} \pi \over L} \cot \Theta_{n (\alpha i)}$;
$\Theta_{nm} =\pi
(y_n-y_m)/L$, $ \Theta_{n(\alpha i)} =\pi
(y_n - x_i^{\alpha})/L$.
The sum can be carried out after expanding
the products and classifying terms by the number of particles involved.
In the end, only the zero particle term and two particle terms are left.
Many particle terms vanish, yielding the following result:
\begin{equation}
{{\hat T}(N)\Phi_G \over \Phi_G} =
-{2\pi t \over L} Q J_h + (2\pi t /L)i \sum_n \sum_{\alpha,i}
\cot \Theta_{n(\alpha i)}.
\end{equation}
This result is valid under the condition $|J_h| \le {L/2} -
[Q+(M_1 + M_2 + \cdots +M_{N-1} )]/2.$

To consider the effects of other parts of the hopping operators, we can
not use the wavefunctions directly, since the hopping will
involve the fermions and holes simultaneously when they operate
on the wavefunctions. We can generalize the idea of spin-rotated version
developed in the recent work of the $1/r^2$ t-J model to this
$SU(N)$ case. For example, to deal with the hopping operator
$T(N-1)=\sum_{i\ne j} c_{i (N-1)}^{\dagger} c_{j (N-1)}$,
we can write the Gutzwiller-Jastrow wavefunctions in terms of the
hole positions and the positions of the fermions of flavors
excluding the $(N-1)$-th flavor. In terms of these coordinates,
the wavefunctions can be found to be still in a similar product form, and
thus effect of the operator $T(N-1)$ can be calculated in the same way
as for $T(N)$.

For the other hopping operators $T(N-2)$, $T(N-3)$, $\cdots, T(2), T(1)$,
similar procedures can be carried on.
After adding all the effects of the hopping operators together, the two
particle
terms vanish since positions of all the fermions and the positions of the holes
span the entire lattice. Thus the Gutzwiller-Jastrow wavefunctions are
found to be exact eigen-states of
the Hamiltonian, with eigen-energies given by
\begin{equation}
E(J_h;J_1, J_2, \cdots, J_{N-1}) = -(2\pi t /L) Q
[ J_h + {\tilde J_h^{(1)}} + { \tilde J_h^{(2)}} + \cdots
+{\tilde J_h^{(N-1)}}],
\end{equation}
where we have ${ \tilde J_h^{(1)}} = J_h -J_{N-1} +L/2$, $
{\tilde J_h^{(2)}} = J_h -J_{N-2} +L/2$, $\cdots$,
${\tilde J_h^{(N-1)}} = J_h -J_1 +L/2$. The many particle terms vanish, and
thus our result holds, under the conditions
\begin{eqnarray}
&&|J_h| \le M_N/2,\nonumber\\
&&|{\tilde J_h^{(1)}}| \le M_{N-1} /2,\nonumber\\
&&| {\tilde J_h^{(2)}}| \le M_{N-2}/2,\nonumber\\
&&\cdots\nonumber\\
&&|{\tilde J_h^{(N-1)}}| \le M_1/2.
\end{eqnarray}
Here the  ground state energy is given by $E_0 = -(2\pi |t|/L)
[ L/2-Q/2 ] Q $.

For this multi-component system, the spectrum can also be written in terms of
more generalized Jastrow functions. Here, we just write down the spectrum
without
getting into the detailed algebra,
\begin{equation}
E = -(2\pi t /L ) \sum_{i=1}^Q K_i + (\pi t Q /L) (L+1),
\end{equation}
where $K_i$ takes values from the region $(1, 2, \cdots, L)$.
Here the result indicates that the spectrum is invariant when changing
the sign of $t$.
Each energy level is determined by a charge configuration such as
$(101010)$ for $Q=3, L=6$, where the ones represent the values occupied
by the charge momenta $K_i$. In this system, the spin
and charge degrees are decoupled
from each other in the entire Hilbert space. On these physical grounds, we
see that for each charge configuration the
degeneracy of the corresponding energy level is given by the number
of the ways to
distribute the free spins among the $L-Q$ empty values.
With this result, we find the free energy per lattice site given by
\begin{equation}
F(T,\mu)/L = - \mu - {T\over 2 \pi}
\int_{-\pi}^{\pi} dq \, \ln [ N + e^{\beta (qt-\mu)}],
\end{equation}
where $\mu$ is the chemical potential of the fermions.
This free energy has also been found to correctly reproduce the first three
terms in the high temperature perturbation expansion.

In summary, we have solved the multi-component Hubbard model
in the strong interaction
limit. In this limit, the spin degrees of freedom decouple
from the charge degrees of freedom in the $\it entire$ excitation spectrum,
and the system is a special example of the
$SU(N)$ Luttinger liquids in the sense of Haldane. We have shown
that the $SU(N)$ Gutzwiller-Jastrow wavefunctions are eigen-states
of the Hamiltonian.

In the end, we notice that in the half filling and large $U$ limit,
our model reduces to the $SU(N)$
Haldane-Shastry spin model with $1/r^2$ exchange interaction.
It seems likely that our multi-component Hubbard model of
the $1/r$ hopping is also completely integrable for $\it arbitrary$
on-site energy $U$ at $\it arbitrary$ filling number\cite{note2}.
However, we have not found any elegant way to
obtain the wavefunctions and the energy spectrum for the finite
on-site energy case. The nested Bethe-ansatz might
be an essential key to the
exact solutions at finite on-site energy. It is also very likely that the
$SU(N)$ system also exhibits a metal-insulator
phase transition at half filling when
changing the bandwidth and the on-site energy, as in
the $SU(2)$ case discovered
by Gebhard and Ruckenstein about two years ago.
It is also of great interest to study the ground
state properties of the system
as a function of the interaction strength, such as, the spin and charge
susceptibilities and various ground state
correlators. It also remains to find the integrability
condition for the model at the finite on-site energy.

This work was supported in part by
the National Science Foundation
under Grant NSF-DMR-89-13692.
The author wishes to thank Prof. P. Coleman for fruitful
discussions and constant encouragment during this work,
and is also grateful to him for reading the manuscript and for useful
comments.

$\dagger$ Address after September 1993: Institut de Physique
Th\'eorique, \'Ecole Polytechnique F\'ed\'erale de Lausanne,
PHB-Ecublens, CH-1015 Lausanne-Switzerland.

\end{document}